\documentclass[epsfig,psfig,aps,twocolumn,prb,showpacs]{revtex4-1}
\usepackage[colorlinks=true,citecolor=red,urlcolor=blue]{hyperref}
\usepackage{amsfonts}
\usepackage{graphicx}
\usepackage{epsfig}
\usepackage{epsf}
\usepackage{amsmath}
\usepackage{physics}
\usepackage{epstopdf}
\usepackage{float} 
\usepackage{amstext}
\newcommand{\be}{\begin{equation}}
\newcommand{\beqn}{\begin{eqnarray}}
\newcommand{\eeq}{\end{equation}}
\newcommand{\ee}{\end{equation}}

\begin{document}

\title{Magnetic Fabry-P\'erot interferometer for valley filtering in a honeycomb-dice model}

\author{F. Bouhadida}
\author{L. Mandhour}
\author{S. Charfi-Kaddour}

\affiliation{Laboratoire de Physique de la Mati\`ere Condens\'ee, D\'epartement de Physique,
Facult\'e des Sciences de Tunis, Universit\'e Tunis El Manar, Campus Universitaire 1060 Tunis, Tunisia}

\begin{abstract}
 Here we theoretically investigate the valley-dependent transmission of particles through a combined electric and magnetic barrier in the $\alpha-T_3$ model which interpolates between the honeycomb and the dice lattices. We put forward that the combination of the Fabry-P\'erot interferences and the magnetic field leads to a perfect transmission for one valley and a suppression of the transmission for the other valley. When only one of the barriers (magnetic or electric) is present, no valley polarized current can be produced. By tuning the Fermi energy, this valley-dependent peculiar  behavior can be used as valley filtering. Our results show that highly efficient valley filtering with maximum conductivity and polarization can be achieved by controlling the value of the magnetic field and the electric barrier width and height.

\end{abstract}

\date{\today}
\maketitle

\section{Introduction}
The \textit{valley} degree of freedom in condensed matter materials opened the door to a new field referred as \textit{valleytronics}\cite{Schaibley,Rycerz} which is reminiscent of the spintronics \cite{Zutic}. The band structure of the valleytronics  materials must have at least two inequivalent valleys in order to control the current coming from each valley.
Graphene, a single sheet of carbon atoms arranged in a honeycomb lattice (HCL) that offers two \textit{valleys} $K$ and $K'$, is one of the most used valleytronics materials. In the existing literature different methods to obtain a valley filter in graphene have been presented such as lattice strain \cite{Yesilyurt, Hung, Zhai, Fujita, Zhao}, line defect \cite{Cheng, Gunlycke}, trigonal warping effect \cite{Pereira}, the effect of mirror-symmetry breaking \cite{Asmar} and strong electrostatic potential \cite{Wang}.
The black phosphorus has also been used as valley filter with the merging of Dirac cones \cite{Ang}.
As well, the presence of a magnetic field and a potential barrier in bilayer graphene acts as a valley filter \cite{Park}.  

Raoux {\it et al.} \cite{Raoux}  introduced the $\alpha-T_3$ model that is obtained from a HCL by adding in the center of each hexagon an extra site related to one of the sites $A$ or $B$ by an $\alpha$-dependent  hopping amplitude that adjusts the coupling between atoms. The interest of the $\alpha-T_3$ model lies in the fact that it interpolates via the parameter $\alpha$ between graphene ($\alpha=0$) and dice lattice (or $T_3$) \cite{Sutherland, Vidal} ($\alpha=1$) and that its energy spectrum does not depend on $\alpha$ . It has been showed that $Hg_{1-x}Cd_xTe$ \cite{Malcolm}  for a critical value $x=0.17$, maps onto the $\alpha-T_3$ structure for a parameter $\alpha = \frac{1}{\sqrt{3}}$. The energy band spectrum of the $\alpha-T_3$ model is composed of two dispersive bands similar to those of graphene and a dispersionless flat band that crosses the $K$ and $K'$ valleys. One of the most striking characteristics for the $\alpha-T_3$ model is that the dispersion relation at low energy is linear and the particles behave as massless Dirac particles with a hybrid pseudospin that is an admixture of pseudospin $S=1/2$ of graphene and  pseudospin $S=1$ of dice lattice. As a consequence, many unusual properties have been attributed to the $\alpha-T_3$ model such as Klein tunneling \cite{Katsnelson} that consists in perfect transmission across a potential barrier that occurs at normal incidence and it is related to the conservation of the pseudospin \cite{Illes2017}. Perfect transmission at oblique incidence can arise from Fabry-P\'erot interferences of the particle bouncing between the two interfaces of the barrier for all values of $\alpha$ \cite{Illes2017}. For the dice lattice, when the energy of the incident particle is equal to half of the barrier height, a peculiar property called super Klein tunneling \cite{Urban} is observed. It corresponds to the perfect transmission of the particles through the barrier for all the incidence angles.
It has been demonstrated that inhomogeneous magnetic field can suppress the Klein tunneling \cite{Li} and  confine particles in graphene \cite{Martino} and dice lattice \cite{Urban}. Another significant property of the $\alpha-T_3$ model is that when a perpendicular magnetic field is applied on the lattice, the energy becomes quantized into Landau levels which are different for the $K$ and $K'$ valleys when $0 <\alpha < 1$.

Recently, due to its exotic properties, the $\alpha-T_3$ model have been considered for valley filtering. Indeed, Hong-Ya Xu {\it et al.} \cite{Xu}  have used a geometric Valley Hall Effect in the $\alpha-T_3$ model as valley filtering. SK Firoz Islam {\it et al.} \cite{Firoz} have showed that the $\alpha-T_3$ model under a magnetic field acts as a valley filter when subjected to weak spatial electric and magnetic modulation.

We propose, in this paper, highly efficient valley filtering in the $\alpha-T_3$ model by using a combined electric and magnetic barrier. The electric barrier taken alone acts as a Fabry-P\'erot interferometer where the particles can be transmitted across the barrier either by Klein tunneling or by Fabry-P\'erot resonances that  are analogous to those encountered in optics and the current crossing the electric barrier is not valley polarized. Adding the magnetic barrier to the electric barrier  suppresses the Klein tunneling and induces valley-dependent Fabry-P\'erot resonances which correspond to the Onsager semiclassical quantization of cyclotron orbits that coincide with the Landau levels. We also show that by considering a magnetic barrier alone, not only there is no valley-dependent transmission but also the transmission probability is almost the same for all the values of the parameter $\alpha$.

The article is organized as follows. In Section \ref{SII}, we present the $\alpha-T_3$ model. Section \ref{SIII} is devoted to calculate the transmission probability, the conductivity and the polarization for the combined electric and magnetic barrier. In Section \ref{SIV}, our results are presented and the valley-dependent transmission is discussed. Finally, Section \ref{SV}, summarizes and concludes the present article.

\section{Presentation of the $\alpha-T_3$ model } \label{SII}

Graphene is a 2D layer of carbon atoms arranged in a HCL which results from the juxtaposition of regular hexagons. The crystallographic structure of graphene is then described by a triangular Bravais lattice whose primitive cell consists of two carbon atoms $A$ and $B$ because they are not equivalent from a crystallographic point of view. 

Starting from the HCL of graphene and adding in the center of each hexagon an atom $C$ connected to one of the two inequivalent sites (for exemple $B$) with a hopping amplitude $t_{BC}=t_{AB}=\frac{t}{\sqrt{2}}$ where $t$ is the hopping amplitude between the sites $A$ and $B$ of HCL, we obtain the dice lattice $T_3$ \cite{Sutherland}.

\begin{figure}[h!]
  \centering
\setlength{\unitlength}{1mm}
\includegraphics[width=0.35\textwidth]{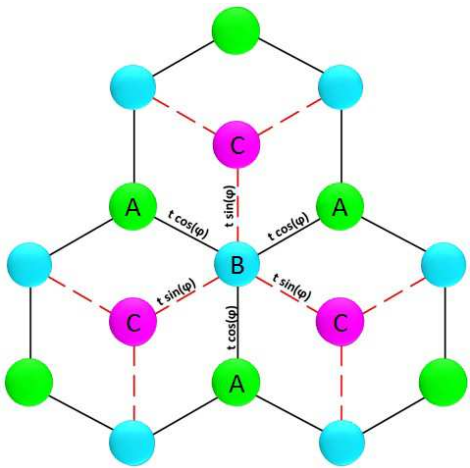}
\caption{(Color online) The $\alpha-T_3$ lattice. There are three sites $A$, $B$ and $C$ in each unit cell. The hopping amplitudes between the sites $A $ and $B$ are denoted $t\cos\varphi$ and those between the sites $B$ and $C$ are given by $t\sin\varphi$. }
\label{figure1}
\end{figure}

The $\alpha-T_3$ model interpolates between HCL and dice lattice via the parameter $\alpha$ such as the hopping amplitudes are given by: $t_{BC}= \alpha t_{AB}$. Indeed, we obtain HCL for $\alpha=0$ and the dice lattice for $\alpha=1$. Fig. \ref{figure1} shows the arrangement of the atoms within the $\alpha-T_3$ model. For computing convenience, we introduce the parameter $\varphi$ such as $\tan\varphi=\alpha$.  

\begin{figure}[h!]
  \centering
\setlength{\unitlength}{1mm}
\includegraphics[width=0.48\textwidth]{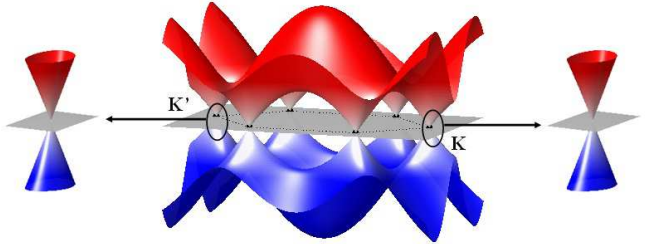}
\caption{(Color online) The energy spectrum of the $\alpha-T_3$ model consists of two dispersive bands and a flat band. The zoom-ins give the low energy spectrum around the $K$ and $K'$ valleys.}
\label{figure2}
\end{figure}

The low energy Hamiltonian for the $\alpha-T_3$ model can be written as:

\be
 H_o= \begin{pmatrix}
0 & f^\chi_o(k)\cos\varphi & 0\\ 
 f_o^{*\chi}(k)\cos\varphi& 0 & f^\chi_o(k)\sin\varphi\\ 
 0& f_o^{*\chi}(k)\sin\varphi  & 0
\end{pmatrix} 
\label{hamilton0} 
\ee
where $f^\chi_o(k)=\hbar v_F(\chi k_x-ik_y)$.  Here $\chi=\pm 1$ is the valley index and $v_F$ is the Fermi velocity. 
The spectrum consists of two cones whose energies are $E_s=s\hbar v_F\left | k \right |$ where $s=\pm$ refers to the band index and a flat band with energy $E=0$.
The corresponding wave functions are given by \cite{Raoux}:
\begin{subequations}
\be
\psi_s(\vec{r})=\frac{1}{\sqrt{2}}\begin{pmatrix}
\cos\varphi e^{i\theta_\chi}\\ 
s\\ 
\sin\varphi e^{-i\theta_\chi}
\end{pmatrix} e^{i\vec{k}\vec{r}}
\label{bands} 
\ee
\be
\psi_0(\vec{r})=\begin{pmatrix}
\sin\varphi e^{i\theta_\chi}\\ 
0\\ 
-\cos\varphi e^{-i\theta_\chi}
\end{pmatrix} e^{i\vec{k}\vec{r}}
\label{band0} 
\ee
\end{subequations}
where $\theta_\chi= \arg{f^\chi_o(k)}$.
The bands of the spectrum touch at the six corners of the Brillouin zone with two inequivalent points corresponding to the $K$ and $K'$ valleys (Fig.  \ref{figure2}).
We see from the expression of the energy that the band structure is independent of the parameter $\alpha$ whereas the eigenfunctions do.

The $\alpha-T_3$ model is characterized by an $\alpha$-dependent Berry phase that is continuously variable from $\pi$ (for graphene) to $0$ (for the dice lattice). The expression of the Berry phase is given respectively for the dispersive bands and for the flat band by  \cite{Raoux}:
\begin{subequations}
\be
\theta_B =\chi \pi \cos2\varphi
\label{berry} 
\ee
\be
\theta_{0B} =-\chi 2\pi \cos2\varphi
\label{berry0} 
\ee
\end{subequations}

When subjected to a magnetic field, the energy becomes quantized, \textit{i.e.} the electronic density is condensed into Landau levels. The energy of the Landau levels for the dispersive bands in the $\alpha-T_3$ model ($0<\alpha<1$) is different for the $K$ and $K'$ valleys and is given by : 
\be
\mathcal{E}_{n}^{\chi}=s\hbar\omega _c\sqrt{n+\frac{1}{2}-\frac{\chi}{2} \cos2\varphi}
\label{LL} 
\ee
where $n=0,1,...$, the cyclotron pulsation is $\omega _c=\sqrt{2}\frac{v_F}{l_B}$ with $l_B=\sqrt{\frac{\hbar}{eB}}$ being the magnetic length.\\ 
The energy of the flat band remains zero.
Note that for the graphene and the dice lattice, the Landau levels at the $K$ and $K'$ points are the same.

\section{Tunneling through an electric and magnetic barrier}\label{SIII}

\begin{figure}[h!]
  \centering
\setlength{\unitlength}{1mm}
\includegraphics[width=0.45\textwidth]{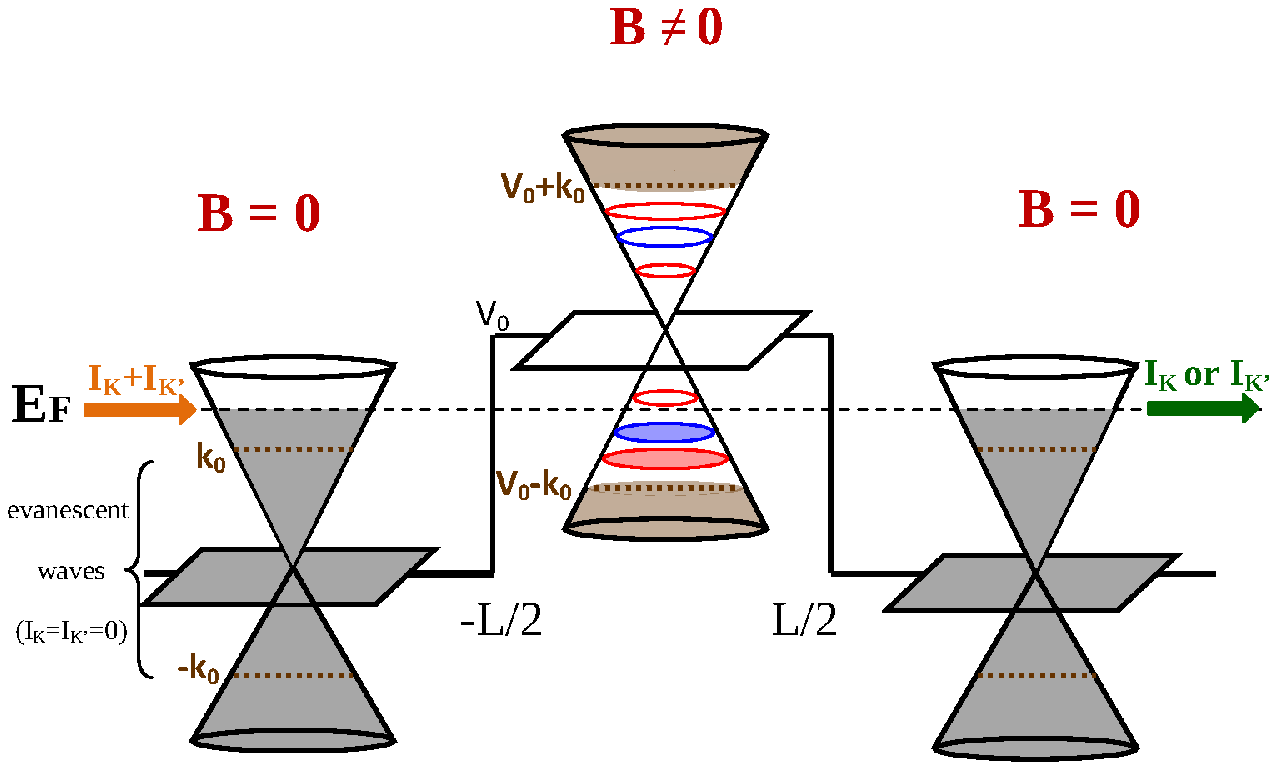}
\includegraphics[width=0.45\textwidth]{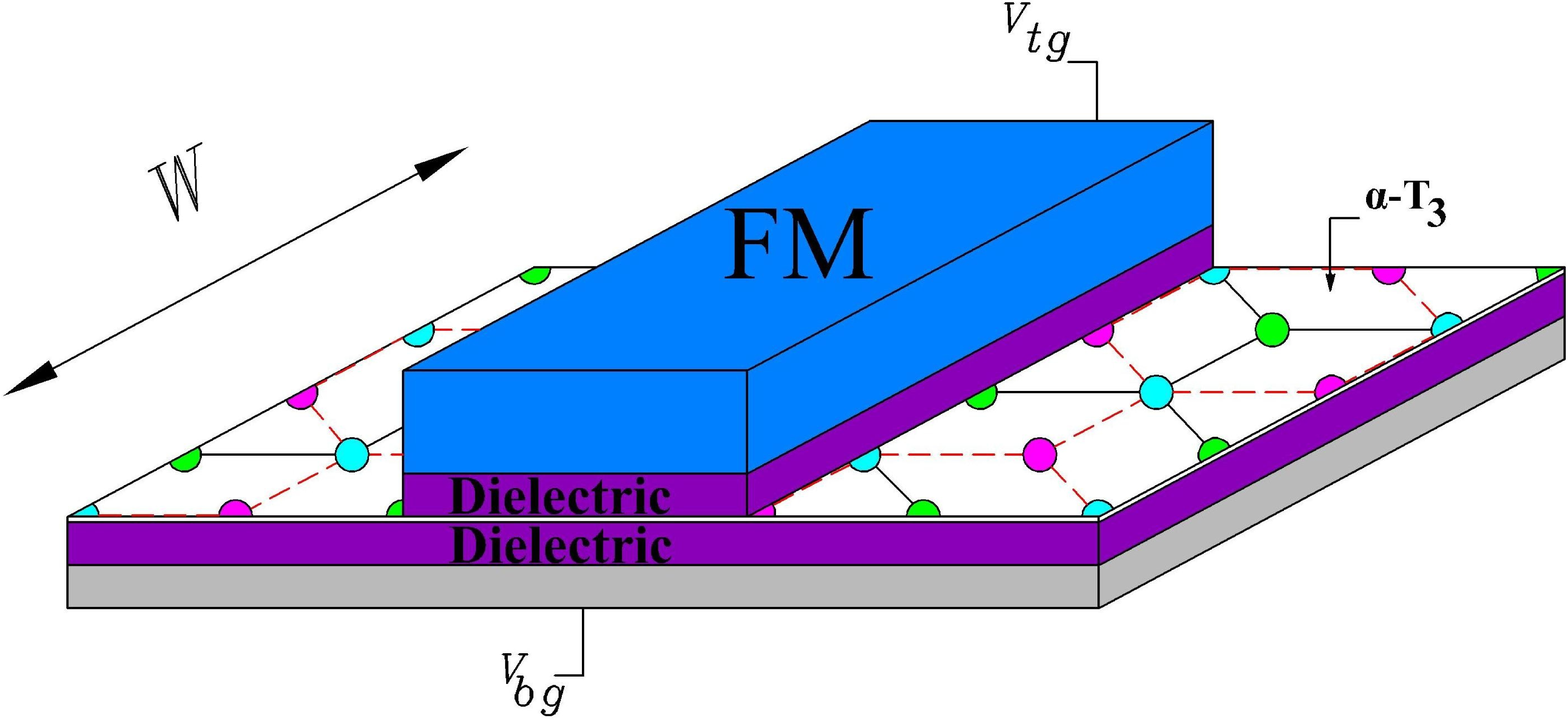}
\caption{ (Color online) Top panel: Schematic of the spectrum in the $\alpha-T_3$ model tunneling through an electric and magnetic barrier. Outside the barrier, the occupied states are depicted in grey. Inside the barrier, the energy is quantized into Landau levels which are different for the $K$ valley (red) and $K'$ valley (blue) when $0<\alpha<1$. The orange and green arrows represent respectively the current injection ($I_K + I_{K'}$) and the valley polarized current ($I_K$ or $I_{K'}$) for energies $E>k_0$. For energies $-k_0<E<k_0$ there is no current due to evanescent waves. Here $k_0$ is given by Eq. (\ref{k0}).\\
Bottom panel:  Schematic of a combined electric and magnetic barrier produced by a ferromagnetic metallic (FM) strip deposited on the central region of the sample, a back gate potential $V_{bg}$ and a top gate potential $V_{tg}$.  $V_{bg}$ and $V_{tg}$ control respectively the Fermi energy $E_F$ and the barrier height. The $\alpha-T_3$ sample is placed between two dielectric layers (purple strips).}
\label{figure_03}
\end{figure}

Here we study the tunneling of Dirac fermions through a combined electric and magnetic barrier. The electric barrier $V(x)$ is of height $V_0$ and width $L$ and the magnetic field $\vec{B}(x)=B(x)\vec{z}$ is perpendicular to the $\alpha-T_3$ lattice (Fig. \ref{figure_03}) such as:
\begin{subequations}
\be 
B(x)=B_0 \Theta (x^{2} -\frac{L^2}{4})
\label{magn barr}
\ee
\be 
V(x)=V_0 \Theta (x^{2} -\frac{L^2}{4})
\label{elec barr}
\ee
\end{subequations}
where $\Theta$ is the Heaviside step function.\\
In this case, the Hamiltonian reads:

\be
 H= \begin{pmatrix}
V(x) & f^\chi(k)\cos\varphi & 0\\ 
 f^{*\chi}(k)\cos\varphi& V(x) & f^\chi(k)\sin\varphi\\ 
 0& f^{*\chi}(k)\sin\varphi  & V(x)
\end{pmatrix} 
\label{hamilton} 
\ee
where $f^\chi(k)=\chi \hbar v_Fk_x-i(\hbar v_Fk_y+ ev_FA_y)$. 
Here $A_y$ is the $y$ component of the vector potential  in the Landau jauge  $\vec{A}=(0,A_y,0)$ and its expression is given by:
\be A_y(x)=B_0\begin{cases}
 &-L/2  \hspace{0.2cm} \text{ if } \hspace{0.5cm}  x<  -L/2\\
 & x  \hspace{0.9cm} \text{ if} \hspace{0.5cm} \left |x  \right |<L/2\\
 &L/2 \hspace{0.5cm}  \text{ if} \hspace{0.6cm}   x>L/2
\end{cases}
\label{pot vect}
\ee\\
The barrier is uniform along the $y$ direction so that the $y$ component of the wave vector is conserved.
The wave function can thus be written as $\Phi \left ( x,y \right )=\Psi \left ( x \right )e^{ik_yy}$ where $\Psi \left ( x \right )=\left ( \psi_A(x),\psi_B(x),\psi_C(x) \right )^T$ is the three-component pseudospinor.\\ 
By solving the Dirac equation $H \Psi(x)=E\Psi(x)$, we obtain the expression of the wave functions in the three regions of the barrier.  
In region I, the wave function is:
\be
\begin{split}
\Psi^\chi_I(x)=&\frac{1}{^{\sqrt{2}}}\begin{pmatrix}
\chi e^{-i\chi \theta} \cos \varphi \\ 
1\\ 
\chi e^{i\chi \theta} \sin \varphi
\end{pmatrix} e^{ik_xx}\\ &-\frac{r_{\chi}}{\sqrt{2}}\begin{pmatrix}
\chi e^{i\chi \theta} \cos \varphi \\ 
-1\\ 
\chi e^{-i\chi \theta} \sin \varphi
\end{pmatrix}e^{-ik_xx}
\label{psi1}
\end{split}
\ee
The first and second terms refer respectively to the incident and reflected waves with the reflexion amplitude $r_{\chi}$ for the valley $\chi$.  
The wave function in region III is given by:
 \be
 \Psi^\chi_{III}(x)=\frac{t_{\chi}}{^{\sqrt{2}}}\begin{pmatrix}
\chi e^{-i\chi \theta'} \cos \varphi \\ 
1\\ 
\chi e^{i\chi \theta'} \sin \varphi
\end{pmatrix} e^{iq_xx}
\label{psi3}
\ee
where $t_{\chi}$ corresponds to the transmission amplitude for the $\chi$ valley. The incidence and the emergence angles are respectively written as: 
\be
\begin{aligned}
&\theta=\arctan (\frac{k_y-k_0}{k_x})  \\ 
&\theta'=\arctan (\frac{k_y+k_0}{q_x})
\label{angles}
\end{aligned}
\ee
The wave vectors in the first and third regions read respectively: 
\be
\begin{aligned}
&k_x=\sqrt{E^2 -(k_y-k_{o})^2}\\
&q_x= \sqrt{E^2 -(k_y+k_{o})^2}
\end{aligned}
\label{kxqx}
\ee
 where
\be 
  k_{o}=\frac{\gamma L}{2} 
 \label{k0}
\ee
 with $\gamma =(\frac{l}{l_B})^2$. We used the following dimensionless variables:
$E \rightarrow\frac{E}{t}$, $V_0 \rightarrow \frac{V_0}{t}$, $k_{xy} \rightarrow k_{xy}l$, $q_x \rightarrow q_xl$, 
$L \rightarrow \frac{L}{l}$ and $x \rightarrow \frac{x}{l}$ with $l=\frac{\hbar v_F}{t}$ and $l_B=\sqrt{\frac{\hbar}{eB}}$ is the magnetic length.

Inside the barrier, the wave functions for the $K$ and $K'$ valleys for $E \neq V_0$  are of the form:
\begin{subequations} 
\be
\begin{split}
 \Psi^K_{II}(x)=&a\begin{pmatrix}
p_{+} D_{p_{+}-1}(X)\cos \varphi \\ 
\lambda D_{p_{+}}(X)\\ 
- D_{p_{+}+1}(X)\sin \varphi
\end{pmatrix} +\\&b\begin{pmatrix}
-p_{+} D_{p_{+}-1}(-X)\cos \varphi \\ 
\lambda D_{p_{+}}(-X)\\ 
  D_{p_{+}+1}(-X) \sin \varphi
\end{pmatrix}
\label{psi2k}
\end{split}
\ee
\be
\begin{split}
 \Psi^{K'}_{II}(x)=&a\begin{pmatrix}
 D_{p_{-}+1}(X)\cos \varphi \\ 
\lambda D_{p_-}(X)\\ 
-p_{-} D_{p_{-}-1}(X)\sin \varphi
\end{pmatrix}+\\&b\begin{pmatrix}
- D_{p_{-}+1}(-X)\cos \varphi \\ 
\lambda D_{p_{-}}(-X)\\ 
 p_{-} D_{p_{-}-1}(-X) \sin \varphi
\end{pmatrix}
\label{psi2kp}
\end{split}
\ee
\end{subequations}

For $E=V_0$ and $ \varphi \neq 0$, \textit{i.e.} at the flat band, the wave functions become:
\begin{subequations} 
\be
\begin{split}
 \Psi^K_{0 II}(x)=&a\begin{pmatrix}
(p_{+}+1) D_{p_{+}-1}(X)\sin \varphi \\ 
0\\ 
 D_{p_{+}+1}(X)\cos \varphi
\end{pmatrix} +\\& b \begin{pmatrix}
(p_{+}+1) D_{p_{+}-1}(-X)\sin \varphi \\ 
0\\ 
  D_{p_{+}+1}(-X) \cos \varphi
\end{pmatrix}
\label{psi02k}
\end{split}
\ee
\be
\begin{split}
 \Psi^{K'}_{0II}(x)=&a\begin{pmatrix}
 D_{p_{-}+1}(X)\sin \varphi \\ 
0\\ 
(p_{-}+1) D_{p_{-}-1}(X)\cos \varphi
\end{pmatrix}+\\&b\begin{pmatrix}
 D_{p_{-}+1}(-X)\sin \varphi \\ 
0\\ 
 (p_-+1) D_{p_{-}-1}(-X) \cos \varphi
\end{pmatrix}
\label{psi02kp}
\end{split}
\ee
\end{subequations}

where $\lambda= \frac{i(E - V_o)}{\sqrt{2 \gamma }}$ and $X=\sqrt{\frac{2}{\gamma}} (k_y + \gamma x) $.
Those latter wave functions are written using the parabolic cylinder functions $D_{p_{\chi}}$ \cite{Gradshteyn} of order $p_{\chi}$ given by:
\be p_{\chi}= \frac{1}{2}\left ( \frac{(E - V_o)^2}{\gamma}+ \chi\cos 2 \varphi -1 \right )
\label{eqp}
\ee

The matching conditions for particles in the $\alpha-T_3$ model \cite{Illes2017} are different from Schr\"odinger and Dirac particles. For that, we integrate the eigenvalue equation $H\Psi \left ( x \right )=E \Psi \left ( x \right )$ over the interval $[x_o-\eta , x_o+\eta]$ by taking $k_x=-i\frac{\partial}{\partial x}$ and we find:
$\left [ S_x^\varphi \Psi(x_0+\eta)- S_x^\varphi \Psi(x_0-\eta)\right ]=\int_{x_0-\eta}^{x_0+\eta} i\chi [(E-V(x))I_3-(k_y+\gamma A(x))S_y^\varphi]$ where $I_3$ is the identity matrix, $S_x^\varphi=\begin{pmatrix}
0 & \cos \varphi & 0\\ 
 \cos \varphi& 0 & \sin \varphi \\ 
 0& \sin \varphi &0 
\end{pmatrix}$ and $S_y^\varphi=i\begin{pmatrix}
0 & -\cos \varphi & 0\\ 
 \cos \varphi& 0 & -\sin \varphi \\ 
 0& \sin \varphi &0 
\end{pmatrix}$ .
Sending $\eta$ to $0$, the second member of this latter equation vanishes, which leads to the matching conditions at $x_0$: 
\be S_x^\varphi \Psi(x_0^+)= S_x^\varphi \Psi(x_0^-)
\label{match} \ee \\
By using these boundary conditions for $x=\pm \frac{L}{2}$, we get the transmission amplitude $t_{\chi}$. Details of the calculation are given in Appendix \ref{A}.\\
To calculate the transmission  probability we need to introduce the probability current\cite{Illes2017}:
\be
\vec{J}=\left\{\begin{matrix}
 J_x=v_F(Re[\psi_B^*(\cos \varphi \psi_A+\sin \varphi \psi_C)])\\ 
J_y=v_F(Im[\psi_B^*(\cos \varphi \psi_A-\sin \varphi \psi_C)])
\end{matrix}\right.
\label{current} \ee 
We finally end up with the transmission probability for the $\chi$ valley:
\be 
T_{\chi}= \frac{\left | J_x^{tr} \right |}{\left |J_x^{inc}  \right |} = \left | \frac{\cos \theta'}{\cos \theta} \right | \left | t_\chi \right |^2
\label{Tr}
\ee\\
where $J_x^{tr}$ and $J_x^{inc}$ are the transmitted and incident components of the current along the $x$ direction which are calculated respectively from the first term of the wave function in region I (Eq. (\ref{psi1})) and the wave function in region III (Eq. (\ref{psi3})).
   
The transmission probability can be appreciated by using measurable quantities such as the conductivity which is given by employing the Landauer-B\"uttiker formula \cite{Blanter}: 
 \be
\sigma_\chi =\frac{L}{W}\frac{2e^2}{\hbar}\sum_{k_y}T_{\chi}(k_y)
\label{cond}
\ee\\
Here $W$ corresponds to the width of the barrier in the $y$ direction. The factor $2$ accounts for the spin degeneracy.

 Finally, to better appreciate the difference between the $K$ and $K'$ valleys, it is judicious to calculate the valley polarization that is determined by:

\be 
P= \frac{\sigma_+-\sigma_-}{\sigma_++\sigma_-}
\label{polz}
\ee

It is important to note that particles can tunnel through the barrier, with a transmission probability given by Eq. (\ref{Tr}), only if the wave vectors $k_x$ and $q_x$ (Eq. (\ref{kxqx})) of the incident and transmitted waves are real which leads to $\left | E \right |> k_0$ and $k_0-E <k_y< E - k_0$  as depicted in the blue zone in Fig. \ref{map}.
However, for $ \left | E \right |< k_0$ particules cannot travel the barrier because at least one of the wave vectors $k_x$ and $q_x$ is imaginary. Particularly, when $E- k_0 <k_y<k_0-E$ and $ \left | E \right |< k_0$ (red zone in Fig. \ref{map}), both wave vectors $q_x$ and $k_x$ are imaginary and in this case the spectrum inside the pure magnetic barrier becomes quantized \cite{Ramezani2010}. 
In order to get the spectrum of the bound states, we consider the Dirac equation $H\Psi(x)=E\Psi(x)$ that leads to the Schr\"odinger equation for the second component $\psi_B(x)$ of the wave function $\Psi(x)$:
\be
-\partial^2_x \psi_B(x)+ U(x)\psi_B(x)=E^2\psi_B(x) 
\label{eqpsib}
\ee
where the effective potential reads:
\be 
U(x)=\begin{cases}
&(k_y-k_o)^2 \hspace{1.9cm} x<-L/2\\ 
&(k_y+\gamma x)^2+U_{\chi} \hspace{0.8cm} \left | x \right |<L/2\\ 
&(k_y+k_o)^2 \hspace{1.9cm} x>L/2
\end{cases}
\label{Ueff}
\ee
with:
\be
U_{\chi}=-\chi \gamma \cos 2\varphi
\label{uchi} 
\ee 
\begin{figure}[h!]
  \centering
\setlength{\unitlength}{1mm}
\includegraphics[width=0.43\textwidth]{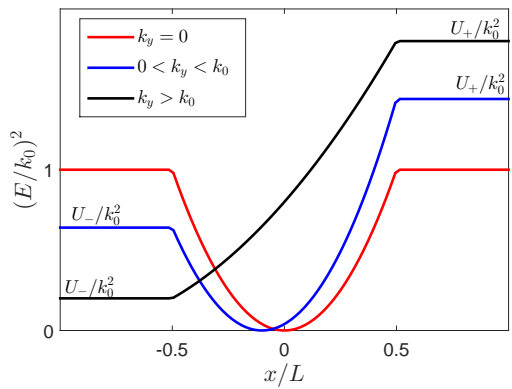}
\caption{ (Color online) Potential $U(x)$ as a function of $x$ for different values of $k_y\geq0$ in the case where $U_{\chi}=0$. For $k_y=0$ (red line): symmetric quantum well of depth $k_0^2$. For $0<k_y<k_0$ (blue line): asymmetric quantum well of depth $U_-=(k_y-k_0)^2$. For $k_y>k_0$ (black line): there are no bound states. The expression of $U_+$ is given by $U_+=(k_y+k_0)^2$.}  
\label{well}
\end{figure}

We plot in Fig. \ref{well}, the potential $U(x)$ as a function of $x$ for different values of $k_y$ in the case where $U_{\chi}=0$. The potential $U(x)$ represents a quantum well for $0 \leq k_y<k_0$ (blue and red lines) and thus the spectrum is quantized when $E- k_0 <k_y<k_0-E$ and $ \left | E \right |< k_0$ (red zone in Fig. \ref{map}). Otherwise, there are no bound states for $k_y\geq k_0$. Note that there is a maximum of bound states in the case of the symmetric quantum well ($k_y=0$).  
If we take into account the term $U_{\chi}$ the spectrum inside the magnetic barrier is shifted upwards (downwards) for the $K'$ ($K$) and this shift is responsible for the lifting of the valley degeneracy of the bound states of the barrier when $0<\alpha<1$. Hence, the spectrum corresponds approximately to the Landau levels \cite{Ramezani2008} given by Eq. (\ref{LL}) as calculated in the Appendix \ref{B}.\\
 The presence of the electric barrier ($V(x)\neq 0$) enables to shift upwards the spectrum by $V_0$ such as the progressive waves find inside the barrier the bound states (see Fig. \ref{figure_03}) which leads to a valley dependent transmission and that is the aim of this work. However, the pure magnetic barrier ($V(x)=0$) is not sufficient to obtain a valley-dependent transmission. Indeed, particles traveling the barrier ($\left | E \right |> k_0$) will find inside the barrier a continuum spectrum that does not enable to have a valley-dependent transmission as depicted in Fig. \ref{map}. 
\begin{figure}[h!]
  \centering
\setlength{\unitlength}{1mm}
\includegraphics[width=0.27\textwidth]{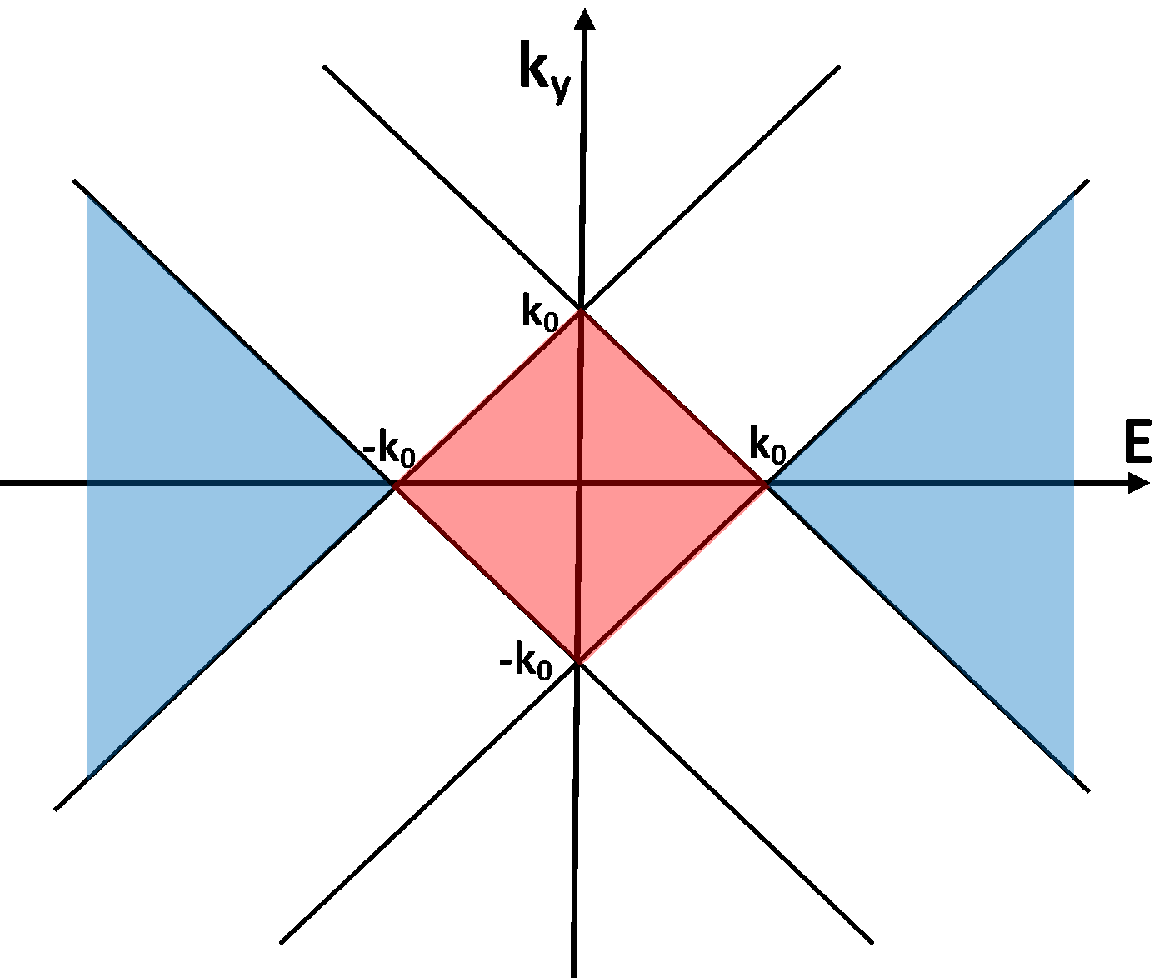}
\includegraphics[width=0.46\textwidth]{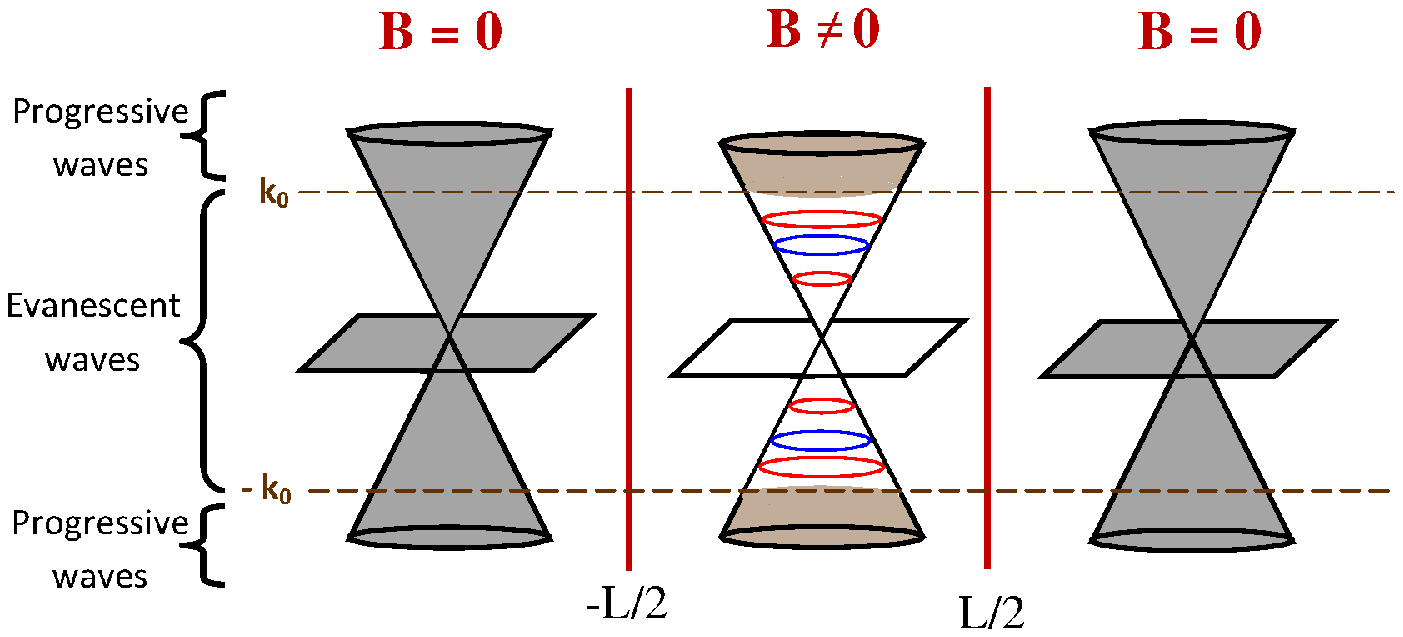}

\caption{ (Color online) Top panel: Schematic representation in the $(k_y,E)$ space of the two zones where the waves are evanescent (red zone with $k_x$ and $q_x$ imaginary) and are progressive (blue zone with $k_x$ and $q_x$ real). \\
Lower panel: Schematic of the spectrum across a pure magnetic barrier in the $\alpha-T_3$ model. For $\left | E \right |< k_0$ no transmission is allowed due to the evanescent waves. For $\left | E \right |> k_0$, the particle can travel the barrier in which there is a continuum spectrum.}
\label{map}
\end{figure}

In order to make an optimal study of the transmission probability, we also take into consideration the potential barrier in the absence of magnetic field. In that case, the wave functions outside the barrier are given from Eq. (\ref{psi1}) and Eq. (\ref{psi3}) by taking the magnetic field $B=0$. Inside the barrier, the wave function for $E\neq V_0$ becomes:

\begin{equation}
\begin{split}
 \Psi_{II}= &\frac{a}{\sqrt{2}} \begin{pmatrix}
e^{-i\Phi}cos\varphi \\ 
s\\ 
e^{i\Phi  }sin\varphi 
\end{pmatrix} e^{iq_0x}\\&+ \frac{b}{\sqrt{2}} \begin{pmatrix}
-e^{i\Phi }cos\varphi \\ 
s\\ 
-e^{-i\Phi  }sin\varphi 
\end{pmatrix}e^{-iq_0x}
\end{split}
\label{eq4b}
\end{equation}
where $\Phi = \arctan\left (\frac{k_y}{q_0} \right)$, $q_0=\sqrt{(E -V_0)^2 -k_y^2}$ and\\ $s=\text{sign} (E-V_0)$.

For $E=V_0$, the Schrodinger equation $H\Psi(x) = E \Psi(x)$ gives:
\be
\begin{cases}
&\cos \varphi (\partial_x+k_y)\psi_B(x)=0\\ 
&\cos \varphi (\partial_x-k_y) \psi_A(x) +\sin \varphi (\partial_x+k_y) \psi_C(x) =0\\ 
&\sin \varphi (\partial_x-k_y) \psi_B(x)=0
\end{cases}
\label{eq04b}
\ee

The solutions depend on either $k_y=0$ or $k_y\neq0$:

(i) for $k_y=0$ and $\varphi \neq 0$ , the wave function is a constant:
$\Psi^{k_y=0}_{0II}(x)=\begin{pmatrix}
a\\ 
b\\ 
c
\end{pmatrix}$.

(ii) for $k_y \neq 0$ and $\varphi \neq 0$, the wave function is given by:

$\Psi^{k_y\neq0}_{0II}(x)=\begin{pmatrix}
a e^{k_yx}\\ 
0\\ 
b e^{-k_yx}
\end{pmatrix}$.\\
We obtain the same wave functions for $0<\alpha\leq1$ as those obtained by Urban \textit{et al.} \cite{Urban} for $\alpha=1$. Hence, the transmission probability is given by:
$\left\{\begin{matrix}
T(k_y=0)=1\\ 
T(k_y\neq0)=0
\end{matrix}\right.$ as depicted in Fig. \ref{fig4}.

By using the matching conditions (Eq. (\ref{match})), we obtain the transmission probability in the absence of a magnetic field which is the same for the two valleys:
\be T_o=\frac{1}{1+\frac{4\gamma _+ \gamma _- }{(\gamma _+-\gamma _-)^2}\sin^2(q_oL)}
\label{trB0} \ee
where 
\be
\gamma_{\pm}=(\cos \theta \mp  s\cos \Phi )^2 +\cos^2 2 \varphi (\sin \theta -s \sin \Phi)^2
\label{gammapm} \ee
Here $\theta$ is given by Eq. (\ref{angles}) in the absence of the magnetic field ($B=0$).
This expression of the transmission probability is in coordination with the result found by  E. Illes \textit{et al.} \cite{Illes2017}.

\section{Results and discussion}\label{SIV}

\begin{figure}[h!]
  \centering
\setlength{\unitlength}{1mm}
\includegraphics[width=0.5\textwidth]{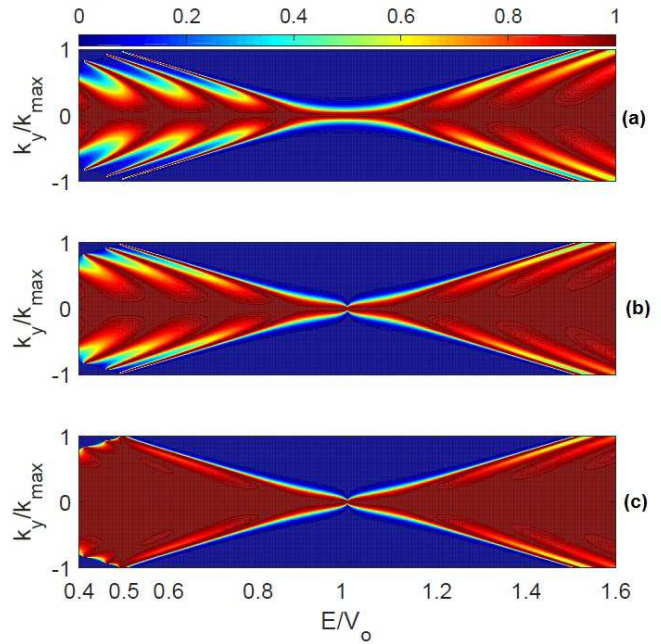}
\caption{(Color online) Transmission probability $T_o$ (Eq. (\ref{trB0})) of the $\alpha-T_3$ model through an electric barrier of width $L=450l$ and height $V_0=0.05t$ when $B=0T$ as a function of the transverse momentum $k_y$ and the energy for different values of the parameter $\alpha$. (a) $\alpha=0$ (HCL), (b) $\alpha=\frac{1}{\sqrt{3}}$ and (c) $\alpha=1$ (dice lattice). The value of $k_y$ must be taken such as the value of $k_x$ (Eq. (\ref{kxqx})) is real: $\left | k_y \right |<k_{max}=E$.}
\label{fig4}
\end{figure}

First of all, we focus on the case of a pure electric barrier. For that, we plot in Fig. \ref{fig4} the transmission probability $T_o$, given by Eq. (\ref{trB0}), as a function of the transverse momentum $k_y$ and the energy $E$ for different values of the parameter $\alpha$. From the expression of the transmission probability given by Eq. (\ref{trB0}), perfect transmission ($T_o=1$) occurs when $\gamma_+ \gamma_-=0$ or $\sin^2(q_0L)=0$. The first case, known as the Klein tunneling effect \cite{Katsnelson}, is available at normal incidence ($k_y=0$) and for all values of the parameter $\alpha$ \cite{Illes2017} and it is related to the conservation of the pseudospin. The second case implies $q_0L=n\pi$ and it corresponds to the Fabry-P\'erot resonances that are analogous to those encountered in optics. 
 Fig. \ref{fig4} (c) also points up a significant behavior for the dice lattice of the transmission probability for an energy $E=\frac{V_0}{2}$: there is a perfect transmission for all values of the incidence angle and the barrier acts as if it was fully transparent and this is the so-called super Klein tunneling\cite{Urban}. Indeed, in this case, the parameter $\gamma_-$ given by Eq. (\ref{gammapm}) vanishes for all the values of $k_y$ leading to a perfect transmission ($T_o=1$).
This figure also spotlights the fact that the transparency of the barrier enhances when the parameter $\alpha$ increases \cite{Illes2017}.
It is important to retain that the transmission probability is exactly the same for the $K$ and $K'$ valleys for all the values of the parameter $\alpha$ in the presence of an electric barrier.  

\begin{figure}[h!]
  \centering
\setlength{\unitlength}{1mm}

\includegraphics[width=0.175\textwidth]{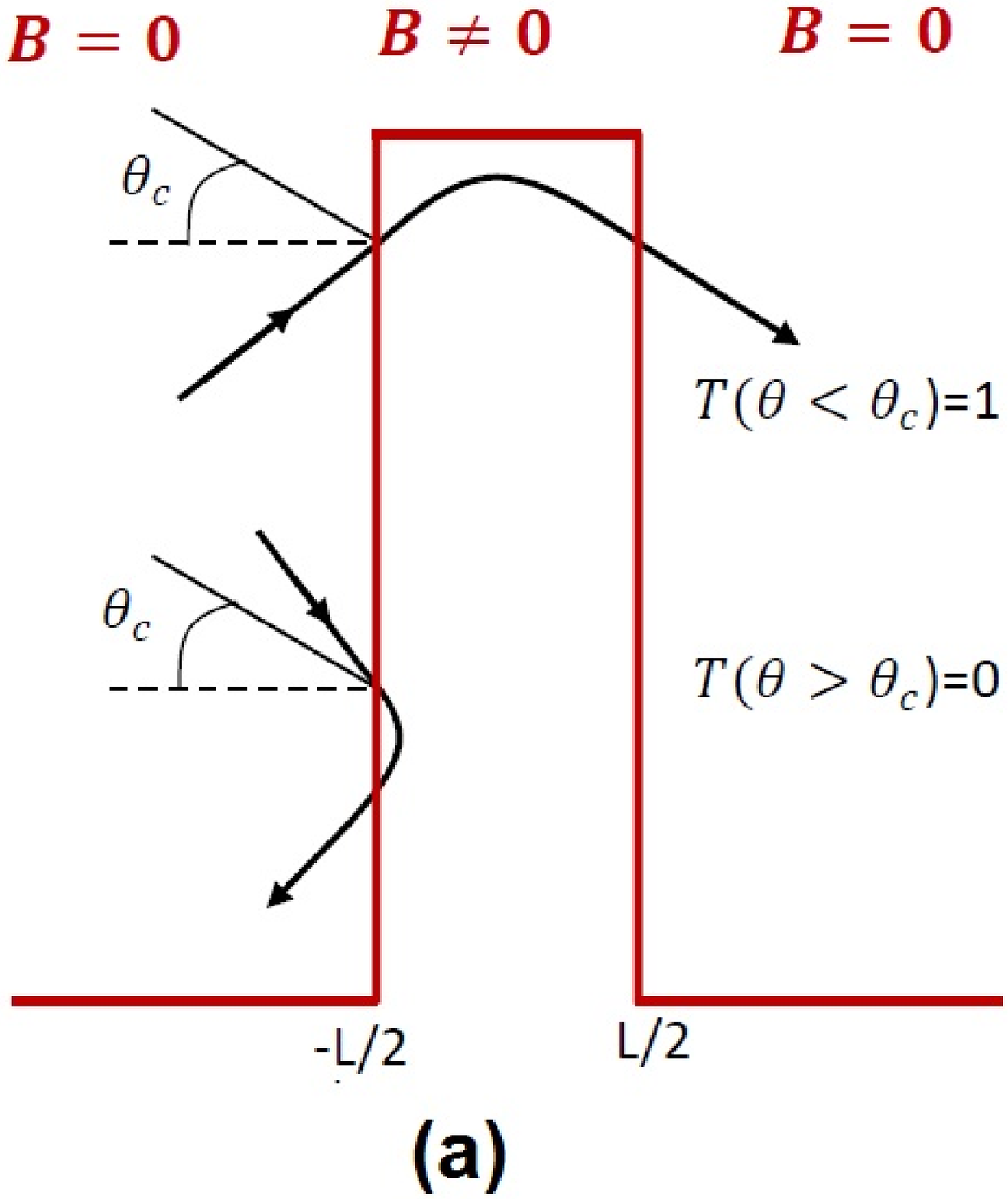}
\includegraphics[width=0.3\textwidth]{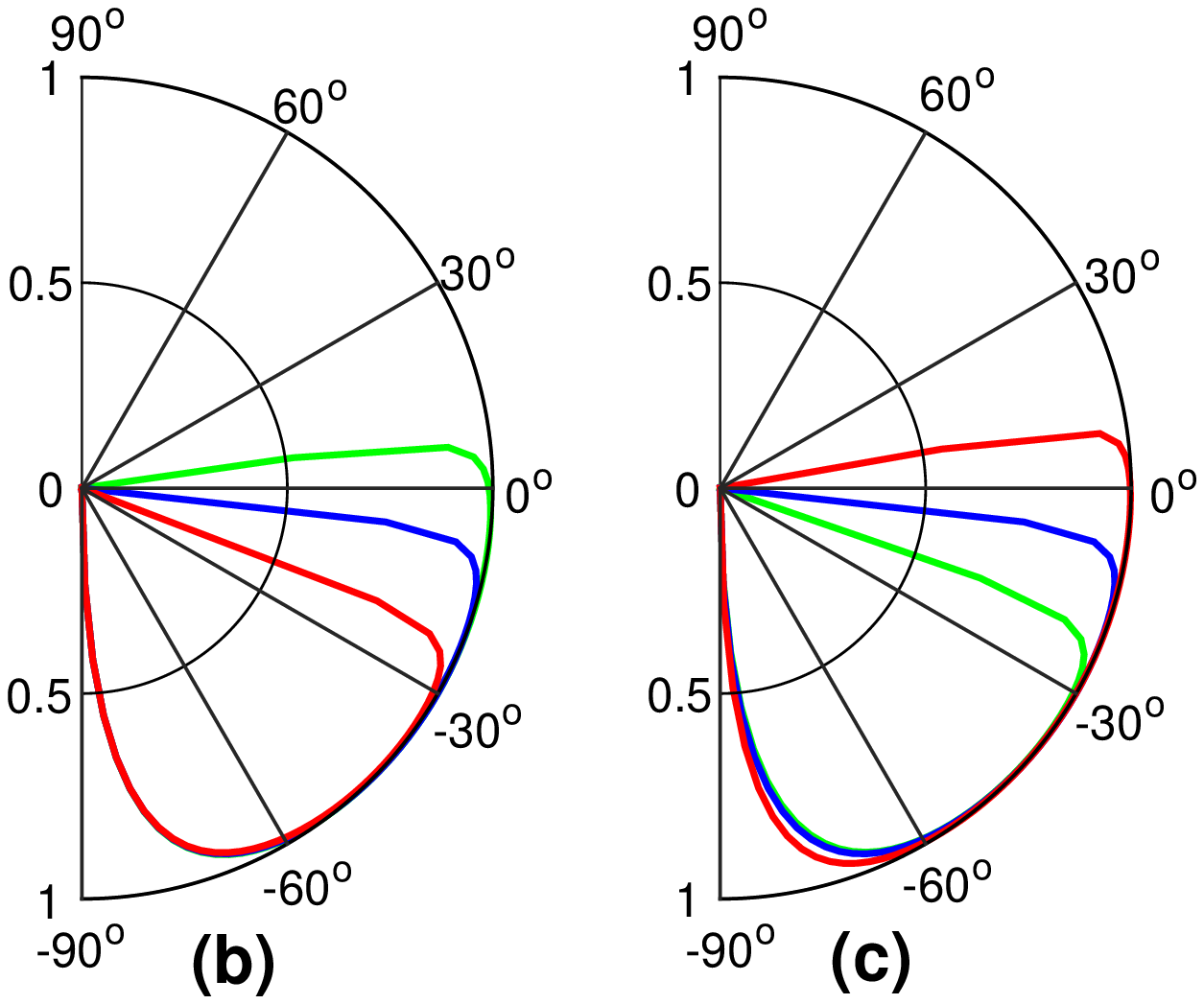}
\caption{(Color online) (a) Illustration of semiclassical trajectories of particles through a magnetic barrier for $\theta<\theta_c$ and $\theta>\theta_c$ where $\theta_c$ is the critical incident angle given by Eq. (\ref{thetac}). (b-c) Polar plot of the transmission probability for a parameter $\alpha=1/\sqrt{3}$ through a magnetic barrier ($B=1T$) as a function of the incidence angle for different values of the barrier width $L$ (b) and different values of the energy $E$ (c). In panel (b), the energy is taken $E=0.03t$ and the green line corresponds to $L=350 l$, the blue line is $L=450 l$ and the red line corresponds to $L=550 l$.
In panel (c), the barrier width is taken $L=450l$, the green line is $E=0.026t$, the blue line is $E=0.03t$ and the red one corresponds to $E=0.04t$.}
\label{fig5}
\end{figure}

Then we take into consideration the tunneling through only a magnetic barrier. For that, we depicted in Fig. \ref{fig5} the polar plot of the transmission probability for a parameter $\alpha=\frac{1}{\sqrt{3}}$ through a magnetic barrier as a function of the incident angle for different values of the barrier width $L$ and the energy $E$. In this case, the transmission probability is given in Appendix \ref{A} in the absence of the electric barrier ($V_o=0$). 

To interpret these results, we use the semiclassical approximation where the particle describes a cyclotron orbit of radius $r_c=\frac{\left | E \right |}{\gamma}$ \cite{Fuchs2010} inside the magnetic barrier. The emergence of the particle at the second interface of the barrier depends on the incident angle $\theta$, the cyclotron radius $r_c$ and the width of the barrier $L$. Using the conservation of the transverse momentum $k_y$ and from Eq. (\ref{angles}), we obtain the relation: $\sin\theta'=\sin \theta +\frac{2k_0}{E}$. We see from this relation that the transmission probability vanishes for all the incidence angles $\theta$  for the energies $\left | E \right | < k_0$ due to evanescent waves as discussed above and depicted in Fig. \ref{map} (red zone).

 In the case where $\left | E \right | > k_0$, the particle can emerge if $-1<\sin \theta +\frac{2k_0}{E}<1$ which implies that:
\be  
\sin\theta<1-\frac{2k_0}{E}= \sin \theta_c 
\label{thetac}
\ee
where $\theta_c$ is the critical incident angle. Consequently, if $\theta<\theta_c$ (blue zone in Fig. \ref{map}), there is a perfect transmission while when $\theta>\theta_c$, the particle makes a full turn and is reflected as illustrated in Fig. \ref{fig5}(a) and hence the magnetic barrier in the $\alpha-T_3$ model confines the Dirac Weyl quasiparticles as in the case of graphene \cite{Martino} and the dice lattice \cite{Urban}. Thus, when considering only a magnetic barrier, the transmission is independent of the parameter $\alpha$ as mentioned by Urban {\it et al.} \cite{Urban} for the limiting cases (graphene and dice lattice). \\ 
We clearly see that for a given energy (Fig. \ref{fig5}(b)) when we increase the width of the magnetic barrier, the critical angle (Eq. (\ref{thetac})) decreases leading to a diminution of the transmission probability. In Fig. \ref{fig5}(c), we notice that for a given $L$ the transmission probability increases when the energy enhances.
We note that the transmission probability is exactly the same for the $K$ and $K'$ valleys in the case of the presence of only a magnetic barrier for all the values of the parameter $\alpha$. 

\begin{figure}[h!]
  \centering
\setlength{\unitlength}{1mm}
\includegraphics[width=0.54\textwidth]{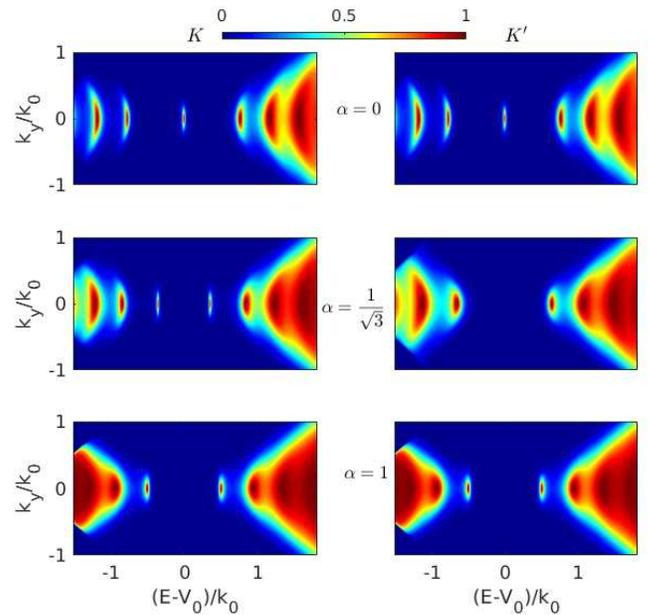} 

\caption{(Color online) Transmission probability $T_{\chi}$ (Eq. (\ref{Tr})) for the two valleys $K$ and $K'$ through a combined electric and magnetic barrier as a function of transverse momentum $k_y$ and the energy for the different values of the parameter $\alpha$ written in the figure.  The parameters used here are:   \\$V_0=0.05t$, $B=1T$ and $L=450l$.}
\label{fig6}
\end{figure}
Let us now discuss the tunneling through a combined electric and magnetic barrier. To that end, we plot in Fig. \ref{fig6} the transmission probability at the two valleys $K$ and $K'$ as a function of the transverse momentum $k_y$ and the energy for different values of the parameter $\alpha$. We notice first of all that the barrier becomes less transparent in the presence of the magnetic field which is related to the fact that the electron exhibits, in a semiclassical picture, a circular orbit inside the barrier and the probability for the particle to reach the second interface of the barrier diminishes \cite{Ramezani2010}. 
Moreover, regardless of the value of the parameter $\alpha$, there is a suppression of the Klein tunneling \textit{i.e.} there is no perfect transmission at normal incidence ($k_y=0$). In the case of the dice lattice ($\alpha=1$), there is no more super Klein tunneling. For the HCL ($\alpha=0$) and the dice lattice ($\alpha=1$), i.e the limiting cases of the $\alpha-T_3$ model, the transmission  probability is identical for the valleys $K$ and $K'$.\\
The main result of this article is seen for $0<\alpha<1$. Indeed, in Fig. \ref{fig6} for $\alpha= \frac{1}{\sqrt{3}} $ the behavior of the transmission probability is different for $K$ and $K'$ which highlights a valley-dependent transmission. 
\begin{figure}[h!]
  \centering
\setlength{\unitlength}{1mm}
\includegraphics[width=0.52\textwidth]{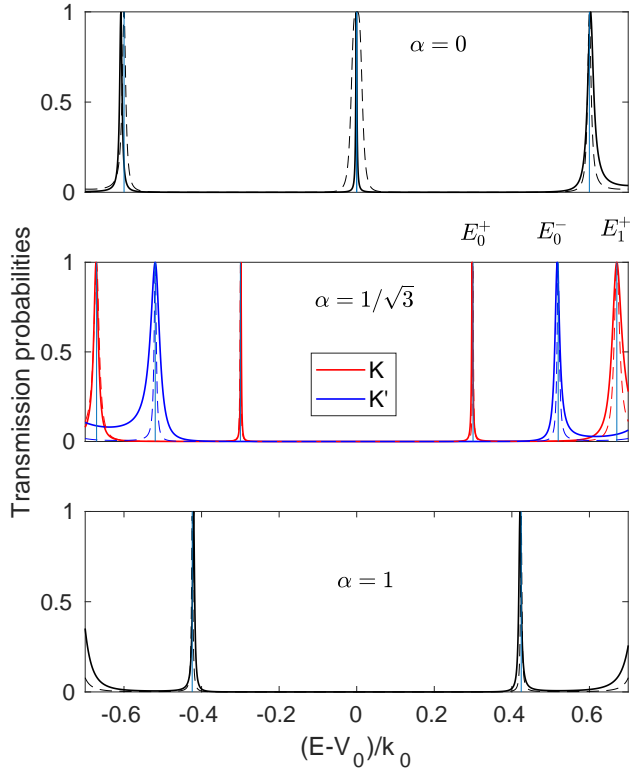} 
\caption{(Color online) Transmission probability $T_{\chi}$ (Eq. (\ref{Tr})) (full lines) and transmission probability $T_{FP}^{\chi}$ (Eq. (\ref{eqFP})) in the Fabry-P\'erot form (dashed lines) at normal incidence ($k_y =0$) through a combined electric and magnetic barrier as a function of the energy for different values of $\alpha$ ($\alpha=0$, $\alpha= \frac{1}{\sqrt{3}} $ and $\alpha=1$ ). The parameters used here are: $V_0=0.05t$, $B=1.5T$ and $L=450l$. The light blue vertical lines represent the Landau levels given by Eq. (\ref{LL}).}
\label{fig7}
\end{figure}
To better understand this behavior, we plot in Fig. \ref{fig7} the transmission probabilities $T_{\chi}$ (full lines) for the two valleys $K$ and $K'$ at $k_y=0$ as a function of the energy  for the three same parameters $\alpha$ as used in Fig. \ref{fig6}. 
For $ \alpha=0$ and $\alpha=1$ (Fig. \ref{fig7}), the transmission probabilities for the two valleys $K$ and $K'$ (plotted in black full lines) are exactly the same which brings out that the particle in those cases present no valley-dependent transmission.
However, for $\alpha=\frac{1}{\sqrt{3}}$ the transmission probabilities for the two valleys are different as illustrated in Fig. \ref{fig7} where the full red line corresponds to the transmission for the $K$ valley and the full blue line refers to the $K'$ valley. 
This last result is also available for all the values of $\alpha$ that verify $0 < \alpha < 1$. This behavior can be understood in the semiclassical picture by the interference between the forward and backward waves inside the barrier. In this case, the longitudinal momentum depends on the $x$ position and is given by $k_x=\sqrt{(E-V_0)^2-\gamma^2 x^2}$ and when the cyclotron radius $r_c=\frac{\left | E-V_0 \right |}{\gamma}< \frac{L}{2}$ we distinguish three regions as depicted in Fig. \ref{figFP}:
\begin{figure}[h!]
  \centering
\setlength{\unitlength}{1mm}
\includegraphics[width=0.48\textwidth]{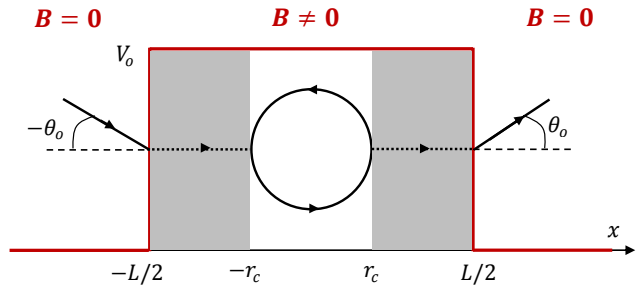}
\caption{(Color online) Illustration of the transmission probability at $k_y=0$ through a combined electric and magnetic barrier in the semiclassical picture. The particles travel the grey regions with evanescent waves while in the middle region ($-r_c<x<r_c$) they perform a cyclotron orbit.   
 The angle $\theta_o=\arcsin  \frac{k_0}{E}$ is calculated from Eq. (\ref{angles}) at $k_y=0$.}
\label{figFP}
\end{figure}

  (i) when $-L/2<x<-r_c$ and $r_c<x<L/2$: these are the classically forbidden regions where $k_x$ is imaginary and the electron is transmitted through these regions with the probabilities in the WKB approximation  \cite{Shytov}: $t=t_1=t_2=\exp\left (-2Im \int_{r_c}^{L/2}k_xdx  \right )$ which is given by the following form: 
  \be
t=\exp\left [-\frac{\left (E-V_0  \right )^2}{\gamma}\left (u\sqrt{u^2-1}-\log \left (u+\sqrt{u^2-1}  \right )  \right)   \right ]
\label{FP} \ee
 where $u=L/(2r_c)$.
  
  (ii) when $-r_c<x<r_c$ the longitudinal momentum $k_x$ is real so the electron exhibits cyclotron orbits with radius $r_c$ and the phase acquired from the interference of the wave scattered between the two classically forbidden region $x=\pm r_c$ is given by:
\be 
 \Delta \theta=\theta_{WKB}+ \theta_1+\theta_2+\theta_B
\label{dteta} \ee
where 
$\theta_{WKB}=2\int_{-r_c}^{r_c}\sqrt{(E-V_o)^2-\gamma^2x^2}dx=\frac{(E-V_o)^2}{\gamma}$,  
$\theta_1+\theta_2 = \pi$ are the backreflection phases for the interfaces 1 and 2 \cite{Shytov} and
$ \theta_{B}$ is the valley-dependent Berry phase given by Eq. (\ref{berry}).

Using the values of the transmission coefficients $t$ (Eq. (\ref{FP})) and the phase $\Delta \theta$ (Eq. (\ref{dteta})), we can write the valley-dependent transmission probability in the Fabry-P\'erot form: 
 
\be
T_{\chi}^{FP}=\frac{t^2}{\left | 1-(1-t)e^{i\Delta \theta } \right |^2}
\label{eqFP} \ee
which is shown in Fig. \ref{fig7} by the dashed lines.\\
We see from this figure that the perfect transmission $T_{\chi}^{FP}=1$ in the Fabry-P\'erot form occurs at the Landau levels $E_n^{\chi}$ (vertical lines). Indeed, this perfect transmission is obtained from the Onsager semiclassical quantization condition \cite{Fuchs2010} $\Delta \theta=2n\pi$ with $n=1,2...$ which is equivalent to Eq. (\ref{LL}) with $\mathcal{E}_n^{\chi}=E_n^{\chi}-V_0$.
The transmission resonances $T_{\chi}=1$ coincide perfectly with the Landau levels $E_n^{\chi}$ showing pronounced peaks when the cyclotron orbit is quite distant from the interfaces of the barrier. Otherwise, the transmission resonances deviate slightly from the Landau levels and become wider. This last result can be attributed to the effect of the walls at $\pm L/2$ on the cyclotron orbit. It changes the energy of the Landau levels in a manner similar to that of the quantum Hall edge channels \cite{Halperin}. For this reason, we focus on the first Landau levels energies $E_0^{\chi}=\mathcal{E}_0^{\chi}+V_0$ (see Eq. (\ref{LL})) at the $K$ and $K'$ valleys where $T_{\chi} \approx T^{\chi}_{FP}$ (see Fig. \ref{fig7} for $\alpha=1/\sqrt{3}$). The transmission probabilities centered around these Landau levels have a Lorentzian distribution and the half width at half maximum is given by:
\be 
\Delta {E_{n}^{\chi }}=\frac{\gamma}{2(E_{n}^{\chi}-V_0)} \frac{t(E_{n}^{\chi}-V_0)}{\sqrt{1-t(E_{n}^{\chi}-V_0)}}
\approx \frac{\sqrt{\frac{\gamma}{2}}}{2\sin\varphi}e^{\frac{-\gamma L^2}{4}}
\label{ene}
\ee\\
This energy width is small when either the magnetic field or the barrier width is large.\\  
Moreover, the value of the parameter $\alpha$ must be chosen such as the overlap between the transmission probabilities around two neighboring Landau levels is minimal. For that, the Landau level $E_0^-$ must be halfway between $E_0^+$ and $E_1^+$ (Fig. \ref{fig7}) which gives $(\cos \varphi - \sin \varphi)= \sqrt{1+\sin^{2}\varphi}-\cos\varphi $. This last result leads to $\varphi=0,57$ which justifies our choice of $\varphi=\frac{\pi}{6}$ ($\alpha=\frac{1}{\sqrt{3}}$). Hence, by choosing the appropriate values of the Fermi energy, the parameter $\alpha$, the magnetic field and the barrier width and height such as the ratio $r_c/(L/2)$ is as small as possible, we can get highly efficient valley-dependent transmission. The simultaneous presence of an electric and  magnetic barrier in the $\alpha-T_3$ model with $0<\alpha<1$ can be thus used as valley filtering.

\begin{figure}[h!] 
  \centering
\setlength{\unitlength}{1mm}
\includegraphics[width=0.5\textwidth]{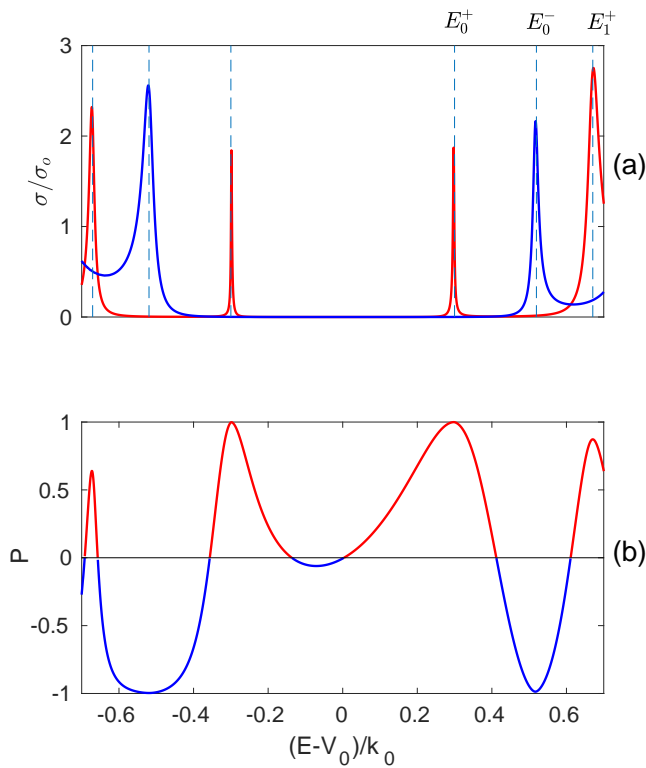} 
\caption{(Color online)(a) Conductivity and (b) polarization for the combined electric and magnetic barrier as a function of the energy for $\alpha=\frac{1}{\sqrt 3}$,  $L=450l$, $B=1.5T$ and $V_o=0.05t$. The vertical light blue dashed lines in Fig. (a) represent the Landau levels given by Eq. (\ref{LL}).}
\label{fig pol1}
\end{figure}
Achieving a valley-dependent conductivity is easily feasible experimentally compared to the transmission probability. For that, we plot in Fig. \ref{fig pol1} the conductivity (Eq. (\ref{cond})) and the polarization (Eq. (\ref{polz})) as a function of the energy for the combined electric and magnetic barrier when $\alpha=\frac{1}{\sqrt{3}}$. In Fig. \ref{fig pol1}(a) the red line corresponds to the conductivity for the $K$ valley and the blue one refers to the $K'$ valley. We also note here that the Landau levels (vertical light blue dashed lines) corresponding to the quantized energy levels coincide perfectly with the peaks of conductivity of the $K$ and $K'$ valleys.
For a value of Fermi energy equal to one of the Landau levels (for example for the $K$ valley), the $K$ component of the electrons pass through the barrier while those of the other valley are reflected. As seen in Fig. \ref{fig pol1}(b), the values taken by the polarization fluctuate between $-1$ and $1$ so that $P=1(-1)$ means that the out-coming current consists of only $K (K')$ contribution.

\begin{figure}[h!]
\setlength{\unitlength}{1mm}
\includegraphics[width=0.5\textwidth]{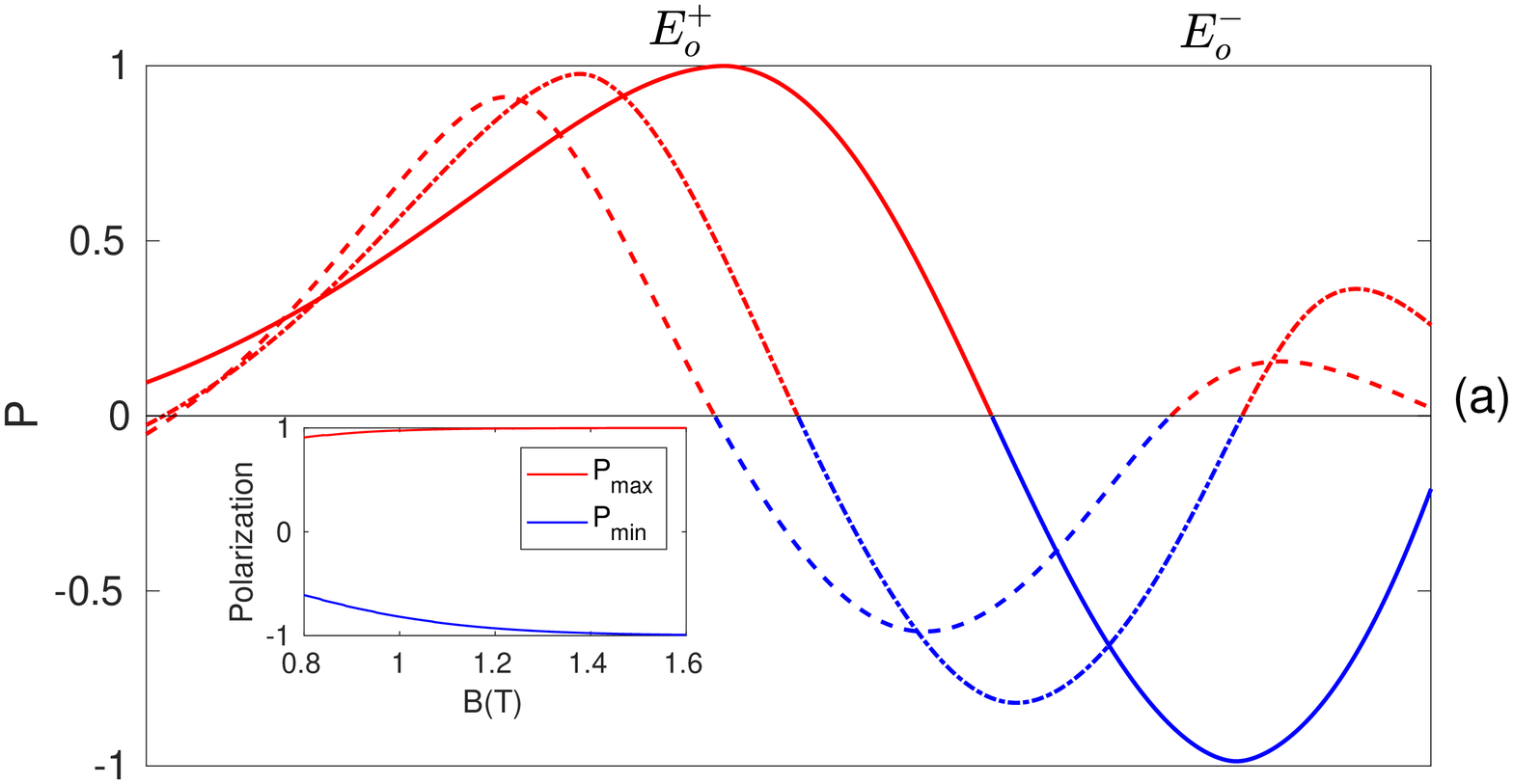} 
\includegraphics[width=0.5\textwidth]{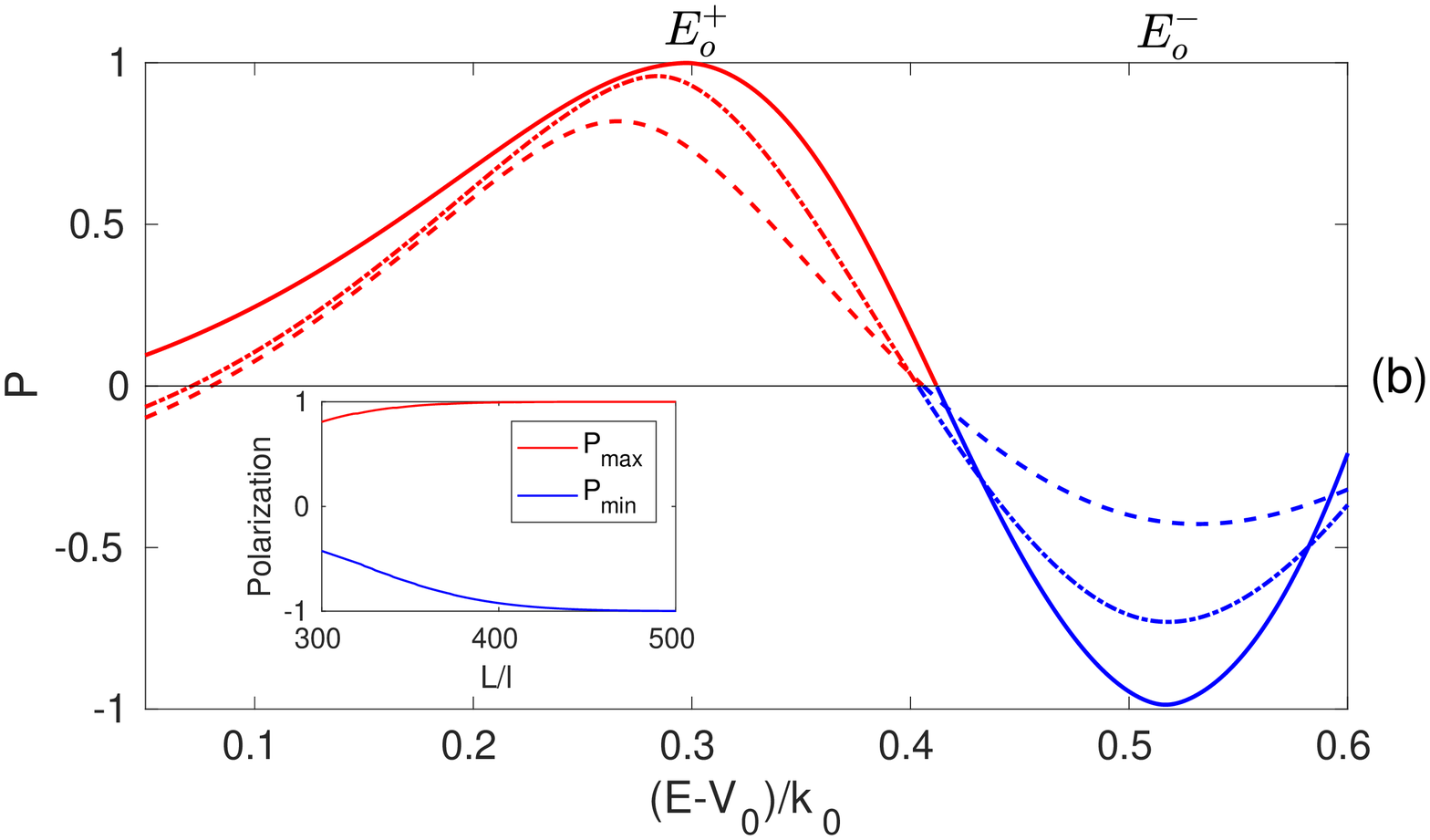} 
\caption{(Color online) Polarization for the combined electric and magnetic barrier as a function of the energy for $\alpha=\frac{1}{\sqrt{3}}$. (a) For $L=450l$, three different values of the magnetic field $B$ were considered: $B=0.8T$ (dashed line), $B=1.0T$ (dotted line) and $B=1.5T$ (full line). (b) For $B=1.5T$, three different values of barrier width were taken: $L=300l$ (dashed line), $L=350l$ (dotted line) and $L=450l$ (full line). The insets represent the maximum and minimum polarizations around the Landau level $E_0^+$ and $E_0^-$ as a function of the magnetic field (a) and the barrier width (b). The value of the parameter $k_0$, given by Eq. (\ref{k0}), is calculated with $B=1.5 T$ and $L=450l$ as in Figs. \ref{fig6}, \ref{fig7} and \ref{fig pol1}. }
\label{fig pol2}
\end{figure}
 Fig. \ref{fig pol2} shows the polarization, when $\alpha=\frac{1}{\sqrt{3}}$, for the electric and magnetic barrier as a function of the energy for different values of the magnetic field (a) and for different values of the barrier width (b). For a given barrier width as in Fig. \ref{fig pol2}(a), the polarization is improved by the enhancement of the magnetic field. The polarization can even become perfect for $B\geq 1.5T$, as seen in the inset, when $L=450l$. Likewise, for a fixed magnetic field $B$, enhancing the width $L$ increases the polarization as seen in Fig. \ref{fig pol2}(b). $L=450l$ is required to reach full polarization as depicted in the inset.  These results can be understood, as discussed previously, by the fact that to achieve a high valley polarization, there must be a minimal overlap between neighbouring Landau levels and the width at half maximum must be  small which is feasible by enhancing $L$ and/or $B$.

\section{Conclusion}\label{SV}

In summary, we studied a way to produce a valley polarized current using a combined magnetic and electric barrier in the $\alpha-T_3$ model. To obtain a valley-dependent transmission, three conditions are necessary:

(i) the magnetic barrier to obtain the Landau levels.

(ii) the electric barrier to shift upwards the spectrum such as the progressives waves find inside the barrier the bound states .

(iii) $0< \alpha <1$ to lift the valley degeneracy between Landau levels.  

Our results showed that perfect transmission ($T=1$) through this kind of barrier occurs at the Landau levels by choosing the appropriate values of the magnetic field and the barrier width and height. In our model, the valley-dependent current traveling the barrier is tuned by the Fermi energy.

As an experimental realization, we propose to use the electric and magnetic barrier on the $Hg_{1-x}Cd_xTe$ at the critical doping $x=0.17$ which maps onto the $\alpha-T_3$ model for a parameter $\alpha=\frac{1}{\sqrt{3}}$ as shown in Fig. \ref{figure_03}. By tuning the Fermi energy controlled by the back gate voltage ($V_{bg}$), we can select the desired current. Such a valley filtering effect could be probed by using a polarizer/analyzer geometry, \textit{i.e.} creating a second such barrier but well separated from the first one in order to avoid interferences between the two barriers.

\section*{Acknowledgments}
We would like to thank J.-N. Fuchs and F. Pi\'echon  for invaluable help and fruitful discussions. We are also indebted to them for a critical reading of the manuscript. This work was partially supported by the Tunisian-French CMCU 15G1306 project.

\appendix 
 \section{The transmission amplitude}
 Applying the matching conditions (Eq. (\ref{match})) at $x =\pm \frac{L}{2}$ for the total wave function given from Eq. (\ref{psi1}), Eq. (\ref{psi3}) and Eqs. (\ref{psi2k}) and (\ref{psi2kp}), we obtain a system of four equations for each valley for $E\neq V_0$:
 \be
\begin{aligned} 
&e^{-ik_x\frac{L}{2}}+r_{\chi} e^{ik_x\frac{L}{2}}=a\lambda D_{p_{\chi}}(X_-)+b\lambda D_{p_{\chi}}(-X_-)\\
&u e^{-ik_x\frac{L}{2}}-r_{\chi} \bar{u}e^{ik_x\frac{L}{2}}= a F^{\chi}(X_-)-b F^{\chi}(-X_-)\\
&t_{\chi} e^{iq_x\frac{L}{2}}=a \lambda D{p_{\chi}}(X_+)+ b \lambda D{p_{\chi}}(-X_+)\\
&vt_{\chi} e^{iq_x\frac{L}{2}}=a F^{\chi}(X_+)- b F^{\chi}(-X_+)
\end{aligned}
\label{sysK}
\ee
Straight forward, the transmission amplitude $t_{\chi}$ for the $K (\chi=1)$ and $K' (\chi=-1)$ valleys  is given by:

\be
t_{\chi}=\frac{C}{x_1 x'_2-x'_1x_2}[x'_2 D_p(X_+)-x'_1 D_p(-X_+)]
\label{tkkp}
\ee
where we have used:
\be
\begin{aligned} 
&C=\lambda(u+\bar{u})e^{-i(k_x-q_x)\frac{L}{2}}\\
&u= \chi \left [\cos^2 \varphi e^{-i\chi\theta}+ \sin^2 \varphi e^{i\chi\theta}  \right ]\\
&v= \chi \left [\cos^2 \varphi e^{-i\chi\theta'}+ \sin^2 \varphi e^{i\chi\theta'}\right ]\\
&x_1= \lambda \bar{u} D_p(X_-)+F^{\chi} (X_-)\\
&x_2= \lambda \bar{u}D_p(-X_-)-F^{\chi} (X_+)\\
&x'_1= \lambda v D_p(X_+)+F^{\chi} (X_+)\\
&x'_2= \lambda v D_p(-X_+)+F^{\chi} (-X_+)\\
&F^+(X)= \cos^2 \varphi p_{+} D_{p_{+}-1}(X) - \sin^2 \varphi D_{p_{+}+1}(X)\\
&F^-(X)= \sin^2 \varphi p_{-} D_{p_{-}-1}(X) - \cos^2 \varphi D_{p_{-}+1}(X)\\
\end{aligned} 
\ee
with $X_{\pm}= \sqrt{\frac{2}{\gamma}} (k_y\pm k_0)$ , $k_0= \frac{\gamma L}{2}$, $\lambda=\frac{i(E-V_0)}{\sqrt{2\gamma}}$, $\theta$ and $\theta'$ are given by Eq. (\ref{angles}) and $p_{\chi}$ by Eq. (\ref{eqp}).
$\bar{u}$ is the conjugate complex  of $u$.

At the flat band in the central region for $E=V_0$ and $\varphi \neq 0$, by applying the matching conditions Eq. (\ref{match})  at $x=\pm \frac{L}{2}$, the total wave functions given from Eqs. (\ref{psi1}), (\ref{psi3}), (\ref{psi02k}) and (\ref{psi02kp}), we obtain a system of four equations for each valley:
\be
\begin{aligned}
&e^{-ik_x\frac{L}{2}}+ r^0_{\chi} e^{ik_x\frac{L}{2}}=0\\
&ue^{-ik_x\frac{L}{2}}-r^0_{\chi} \bar{u} e^{ik_x\frac{L}{2}}= \frac{\sin 2\varphi}{2}  [a G^{\chi}(X_-)+b G^{\chi}(-X_-)]\\
&t^0_{\chi} e^{-iq_x\frac{L}{2}}=0\\
&v t^0_{\chi} e^{-iq_x\frac{L}{2}}= \frac{\sin 2\varphi}{2} [a G^{\chi}(X_+)+b G^{\chi}(-X_+)]\\
\end{aligned}
\ee
where $G^{\chi}(X)=(p_{\chi}+1)D_{p_{\chi}-1}(X)+D_{p_{\chi}+1}(X)$.\\
Hence, the transmission amplitude is $t^0_{\chi}=0$.

\label{A}
 
 \section{Bound states in the magnetic barrier}
 
\begin{figure}[h!]
\setlength{\unitlength}{1mm}
\includegraphics[width=0.5\textwidth]{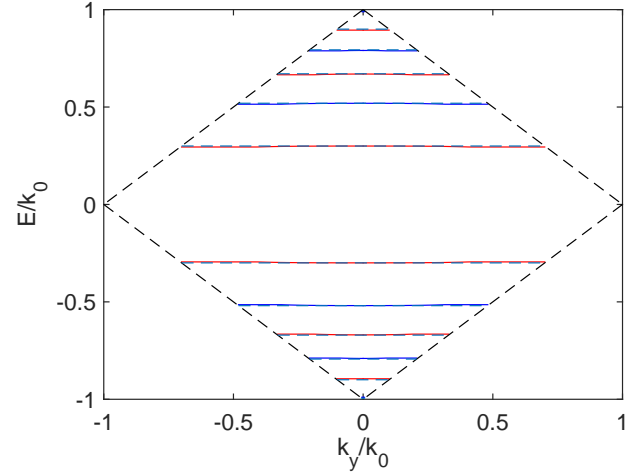}
\caption{(Color online) Energies of bound states as a function of $k_y$ for $\alpha=\frac{1}{\sqrt{3}}$, $L=450l$ and $B=1.5T$. The red and blue continuous lines correspond respectively to the energies of the K and K' valleys. The horizontal light blue dashed lines represent the Landau levels given by Eq.  (\ref{LL}).} 
\label{fig bound}
\end{figure}

From the Schr\"odinger equation (Eq. (\ref{eqpsib})), the wave function $\psi_B(x)$ in the three regions are given by:
\be
\psi_B(x)=
\begin{cases}
c e^{k_xx}, \hspace{3cm} x<-L/2\\ 
aD_{p_{\chi}}(X)+bD_{-1-p_{\chi}}(iX), \hspace{0.01cm}  \left| x \right |<L/2\\ 
d e^{-q_xx}, \hspace{3cm} x>L/2
\end{cases}
\ee 
where $k_x=\sqrt{(k_y-k_0)^2-E^2}$, $q_x=\sqrt{(k_y+k_0)^2-E^2}$, $X= \sqrt{\frac{2}{\gamma}} (k_y+\gamma x)$
and $p_{\chi}$ is given by Eq. (\ref{eqp}).
The continuity of the function $\psi_B(x)$ and its derivative at $x=\pm L/2$
leads to a system of four equations for each valley:
 \be
\begin{aligned} 
&c e^{k_{x}\frac{L}{2}}=a D_{p_{\chi}}(X_-)+b D_{-1-p_{\chi}}(iX_-)\\
&c k_x e^{k_{x}\frac{L}{2}}= a \partial_x D_{p_{\chi}}(X_-)+b \partial_x D_{-1-p_{\chi}}(iX_-)\\
&d e^{-q_{x}\frac{L}{2}}=a D_{p_{\chi}}(X_+)+b D_{-1-p_{\chi}}(iX_+)\\
&-d q_x e^{-q_{x}\frac{L}{2}}= a \partial_x D_{p_{\chi}}(X_+)+b \partial_x D_{-1-p_{\chi}}(iX_+)
\end{aligned}
\label{sysK}
\ee
where $X_{\pm}=\sqrt{\frac{2}{\gamma}} (k_y\pm k_0)$.
By setting to $0$ the determinant of this system, we get the energies of the bound states inside the barrier for $\left | E \right |<k_0$ and $-k_0+E<k_y<k_0-E$ as depicted in Fig. \ref{fig bound} as a function of the transverse momentum $k_y$. 

\vspace{1cm}
$^{\ast}$ Electronic address: lassaad.mandhour@istmt.utm.tn

\label{B}

\end{document}